\documentclass{aa}
\usepackage{graphicx}
\usepackage{deluxetable}
\usepackage{url}
\usepackage{natbib}
\usepackage{rotating}
\usepackage{amsmath}
\usepackage{txfonts}
\begin{document}

   \title{High Resolution Rapid Response observations of compact radio sources with the Ceduna Hobart Interferometer (CHI)}

   \author{Jay M. Blanchard\inst{\ref{tassie},\ref{atnf}}
          \and
          James E. J. Lovell\inst{\ref{tassie}}
          \and
          Roopesh Ojha\inst{\ref{goddard},\ref{catholic}}
          \and
          Matthias Kadler\inst{\ref{wurzburg},\ref{remeis},\ref{cresst},\ref{usra}}
          \and
          John M. Dickey\inst{\ref{tassie}}
          \and
          Philip G. Edwards\inst{\ref{atnf}}}

   \institute{School of Mathematics \& Physics, Private Bag 37, University of Tasmania, Hobart TAS 7001, Australia\label{tassie}
           \and
           Australia Telescope National Facility, CSIRO Astronomy and Space Science, PO Box 76, Epping, NSW 1710, Australia\label{atnf}
           \and
           NASA, Goddard Space Flight Center, Greenbelt, MD 20771, USA\label{goddard}
           \and
	  Institute for Astrophysics \& Computational Sciences (IACS), Dept. of Physics, The Catholic University of America, 620 Michigan Ave., N.E., Washington, DC 20064, USA\label{catholic}
	  \and
           Institut f\"ur Theoretische Physik und Astrophysik, Universit\"at W\"urzburg, Am Hubland, 97074 W\"urzburg, Germany\label{wurzburg}
	  \and
	 Dr. Remeis Sternwarte \& ECAP, Universit\"at Erlangen-N\"urnberg, Sternwartstrasse 7, 96049 Bamberg, Germany\label{remeis}
           \and 
             CRESST/NASA Goddard Space Flight Center, Greenbelt, MD 20771, USA\label{cresst}
           \and 
             Universities Space Research Association, 10211 Wincopin Circle, Suite 500 Columbia, MD 
             21044, USA\label{usra}
             }


 
  \abstract
   {Frequent, simultaneous observations across the electromagnetic 
   spectrum are essential to the study of a 
   range of astrophysical phenomena including Active Galactic Nuclei. 
   A key tool of such studies is the ability to observe an object when it 
   flares i.e. exhibits a rapid and significant increase in its flux density.}
   {We describe the specific observational procedures and the calibration 
   techniques that have been developed and tested to create a single 
   baseline radio interferometer that can rapidly observe a flaring object. This 
   is the \textsl{only} facility that is dedicated to rapid high resolution radio observations 
   of an object south of  $-30$ degrees declination. An immediate application is to 
   provide rapid contemporaneous radio coverage of AGN flaring at $\gamma$-ray 
   frequencies detected by the \textsl{Fermi Gamma-ray Space Telescope}.}
   {A single baseline interferometer was formed with radio telescopes in Hobart, 
   Tasmania and Ceduna, South Australia. A software correlator was set up at 
   the University of Tasmania to correlate these data. }
   {Measurements of the flux densities of flaring objects can be made using our observing 
   strategy within half an hour of a triggering event. These observations can be calibrated with 
   amplitude errors better than 15\%. Lower limits to the brightness temperatures of the sources can also be calculated using CHI.}
{}


  \authorrunning{Blanchard et al.}
  \titlerunning{Ceduna Hobart Interferometer (CHI)}
  \maketitle
%

\section{Introduction\label{sec:intro}}

The launch of the  \textsl{Fermi Gamma-ray Space Telescope} \citep[formerly GLAST;] [] {Atwood2009} on June 11th, 2008, 
has ushered in an era when it is possible to observe astronomical objects, including AGN, simultaneously across 
the entire electromagnetic spectrum. Such multiwavelength observations are essential to understand the behaviour 
of astronomical objects in general and AGN in particular. This motivated the TANAMI program \citep[Tracking Active 
Galactic Nuclei with Austral 
Milliarcsecond Interferometry;] [] {Ojha2010} which provides radio monitoring of $\gamma$-ray loud sources (and a control 
sample) south of  $-30$ degrees declination.  

TANAMI observations are made at two radio frequencies using the telescopes of the Australian Long Baseline 
Array \citep[LBA; e.g.] [] {Ojha2004} augmented by telescopes in South Africa, Chile, Antarctica and New Zealand. This array allows the imaging of the southern sky at milliarcsecond scale or better resolution on a regular basis (e.g., \citealp{Mueller2011}). This observational 
  technique, called Very Long Baseline Interferometry (VLBI), is essentially the only way to measure intrinsic parameters 
  of the jets seen in AGN, as multi-epoch VLBI observations provide the sole direct measurements of their relativistic motion 
  \citep{Cohen2007}. They also play a crucial role in the identification of the nature and location of regions where $\gamma$-ray emission originates in AGN \citep{Jorstad2001,Agudo2011}. TANAMI VLBI observations are supported by radio monitoring programs at arc-second 
  resolution using the Australia Telescope Compact Array (ATCA) and single-dish resolution using the Ceduna 
  radio telescope \citep{McCulloch2005}.  With its associated optical/UV and X-ray programs
and its unique VLBI dual-frequency characteristics, TANAMI has become
one of the major multiwavelength resources for the \textsl{Fermi} mission and
the only one covering sources south of $-30$ degrees.
TANAMI VLBI observations are constrained, however, by the availability of the LBA which only observes in approximately one week blocks every 2$\sim$3 months. This makes it impossible to quickly observe a southern source that exhibits interesting 
behaviour such as rapid changes in flux density at one or more wavebands. 

The Ceduna Hobart Interferometer (CHI) was developed to provide rapid, high resolution observations of sources 
in the southern sky. A fundamental characteristic of AGN is their variability at all wavelengths in which 
they are detected \citep{Ulrich1997,Abdo2010}. Thus a proper study of the physics of AGN requires simultaneous multiwavelength observations, particularly when the flux density of an object changes significantly over a short period of time.
Here we describe the specific observational procedures and calibration 
techniques that have been developed to create this uniquely useful single-baseline interferometer.  It is important to note 
that, even though the development of CHI was motivated by the \textsl{Fermi} and TANAMI programs, it is by no means restricted to 
observations of $\gamma$-ray loud AGN but can be used for multiwavelength study of a range of compact sources e.g. 
optically flaring AGN, X-ray sources being studied by the INTEGRAL and Swift satellites, TeV sources detected by H.E.S.S., radio supernovae and some microquasars.


In this paper, we describe the CHI instrument in 
(Sect.~\ref{sec:instrument}) and describe our observing strategy (Sect.~\ref{sec:observations}). We then explain our calibration procedures (Sect.~\ref{sec:calibration}) and present some results 
in Sect.~\ref{sec:results}. We end with a summary and description 
(Sect.~\ref{sec:summary}) of further development planned for CHI.  
Throughout the paper we use the
cosmology $H_0$=73\,km\,s$^{-1}$\,Mpc$^{-1}$, $\Omega_{m}$=0.27,
$\Omega_{\Lambda}$=0.73 where the symbols have their traditional
meanings. 

\section{CHI: The Ceduna Hobart Interferometer \label{sec:instrument}} 

The Ceduna Hobart Interferometer is a single baseline interferometer formed by a 26 meter antenna located at Mount 
Pleasant, Hobart and a 30 meter antenna located at Ceduna in South Australia. They are both re-purposed antennas that 
are now operated by the Radio Astronomy group at the University of Tasmania. CHI has a physical baseline length of 1704 km corresponding to a resolution of 6.6 milliarcseconds at its current operating frequency of 6.7\,GHz. The Hobart antenna is at a latitude $-$42.8 degrees while the Ceduna antenna is at latitude $-$31.87. Thus CHI can, in principle, observe sources up to mid-northern latitudes and can easily observe the entire southern hemisphere. 

The 30m telescope at Ceduna is a former Telstra Satellite Earth Station, that was converted to astronomical use after its acquisition by the University of Tasmania in 1995. Ceduna uses a standard alt-azimuth mount, with a slew rate of 40 degrees per minute on each axis and an elevation limit of 10 degrees. It is a critical component of the Australian LBA, providing the major east-west baselines. However, Ceduna is primarily used for single dish programs monitoring $\gamma$-ray bright AGN in support of the 
TANAMI program as well as Intra Day Variable \citep[IDV;] [] {Wagner1995} AGN in support of the MASIV \citep[Microarcsecond Scintillation-Induced Variability;] [] {Lovell2003, Lovell2008} program. Ceduna is capable of observations at 2.2, 4.8, 6.7, 8.4, 12.2 and 22 GHz using room temperature receivers.  Since the peak variability in amplitude of scintillating AGN is expected to be close to a frequency of 6.7\,GHz, IDV observations are carried out in this band. As receiver changes at Ceduna are not automated, the standard observing frequency for CHI was selected to be 6.7\,GHz, allowing for a rapid change between observing programs. Despite being uncooled, the 6.7 GHz receiver has a System Equivalent Flux Density (SEFD) of roughly 820 Jy, equivalent to that at Hobart. This is mostly due to the larger collecting area of the 30m dish, as well as the focus type. Hobart is a prime focus instrument with more spillover than the tertiary system at Ceduna.

The 26m telescope at Mount Pleasant was opened on 13 May 1986 after having been moved from Orroral Valley in the Australian Capital Territory where it had been a NASA tracking station for about two decades. It also is a component of the Australian LBA and a vital geodetic antenna regularly participating in International VLBI Service \citep[IVS;] [] {Schluter2007} observations. Hobart uses a non traditional X-Y mount slewing at 40 degrees per minute on both axes to an elevation limit of 7 degrees (apart from a 17 degree limit in the keyhole) and can observe at all major astronomical frequencies from 1.4 through 22\,GHz. The 6.7 GHz receiver at Hobart is cooled to 20 K using helium cryogenics, giving a system equivalent flux density of approximately 820 Jy. The backend system at both telescopes produces two channels (right and left hand circular polarisation) of 32 MHz bandwidth. The baseline sensitivity of a 2 element interferometer is given by equation \ref{eqn:sens} \citep{thompson2001}.

\begin{equation}\label{eqn:sens}{\Delta S_{ij}=\frac{1}{\eta_s}\sqrt{\frac{SEFD_i\times SEFD_j}{2\Delta t \Delta \nu}}}\end{equation}

where $\eta_s$ is the system efficiency (roughly 0.88 for 2 bit sampling), $SEFD$ is the system equivalent flux density in Jy of each antenna, $\Delta t$ is the integration time in seconds and $\Delta \nu$ is the bandwidth in MHz. 

Thus, in just 60 seconds of integration time, CHI can reach a baseline sensitivity of about 15\,mJy which is sufficient to observe all the sources in the TANAMI program. 
The coherence time at 6.7 GHz is approximately 4.6 minutes (calculated using the methods outlined in \citealt{Briggs1983} and checked against the given values in \citealt{Walker1995}). This gives a maximum baseline sensitivity of approximately 7 mJy without phase referencing, more than enough to detect the above mentioned sources. Table~\ref{table:brightnesstemp} lists the brightness temperature sensitivity of CHI to a hypothetical 1 Jansky source for a range of parameters.

CHI data are recorded to disk at 256 Mbps at 2 bit sampling with the LBA Data Recorder system\citep{Phillips2009} and the Ceduna disks are shipped to Hobart for correlation. The data are correlated using an installation of the DiFX software correlator \citep{Deller2007}. The correlator itself consists of a cluster of 6 PCs each containing dual quad-core Xeon processors and 3 GB of memory. The correlator can manage data rates of up to 512 Mbps on a single baseline, giving a 2 times real time correlation speed for a typical CHI observation. The biggest constraint on the speed of analysis of data from CHI however is the time taken for disks from Ceduna to be shipped to the correlator in Hobart. This means a typical turn around time of around 5 days for data to be fully reduced. Transfer over network is not possible due to the limited network capacity to Ceduna and the volume of data recorded.
\section{Observing Strategy for flaring AGN\label{sec:observations}}

Due to the peculiar calibration needs of a single baseline interferometer (see section \ref{sec:calibration}) and the desire to obtain good absolute amplitude calibration, we have developed a non-standard observing strategy. A typical CHI observation runs for twelve hours. Each hour consists of a roughly 50 minute long scan on the target source and a ten minute scan of PKS B193$-$638, the primary amplitude calibrator of the ATCA. PKS B1934$-$638 is observed as traditional amplitude calibration is not possible with a single baseline (see section \ref{sec:calibration} for a discussion).

Observations are usually triggered following an increase in gamma-ray activity as found by the \textsl{Fermi} LAT instrument. The flaring source is also added to the single dish monitoring program, providing a high cadence lightcurve at the observing frequency. If the source is observed to be flaring at radio frequencies multiple CHI observations are then undertaken to provide information on the state and evolution of the flare.

\section{Calibration\label{sec:calibration}}

Calibration of an interferometer with just two elements presents several interesting challenges. Since a minimum of 3 antennas is required for phase closure and 4 antennas for amplitude closure, it is not possible to use the standard VLBI tools of hybrid mapping \citep{Readhead1980} or self-calibration \citep{Pearson1984}.  Thus non-standard amplitude calibration must be performed on CHI data. Calibration requires two stages, a once off initial calibration and a calibration applied to each individual CHI observation.

\begin{figure}[t]
\begin{center}
\leavevmode
\includegraphics[width=0.4\textwidth]{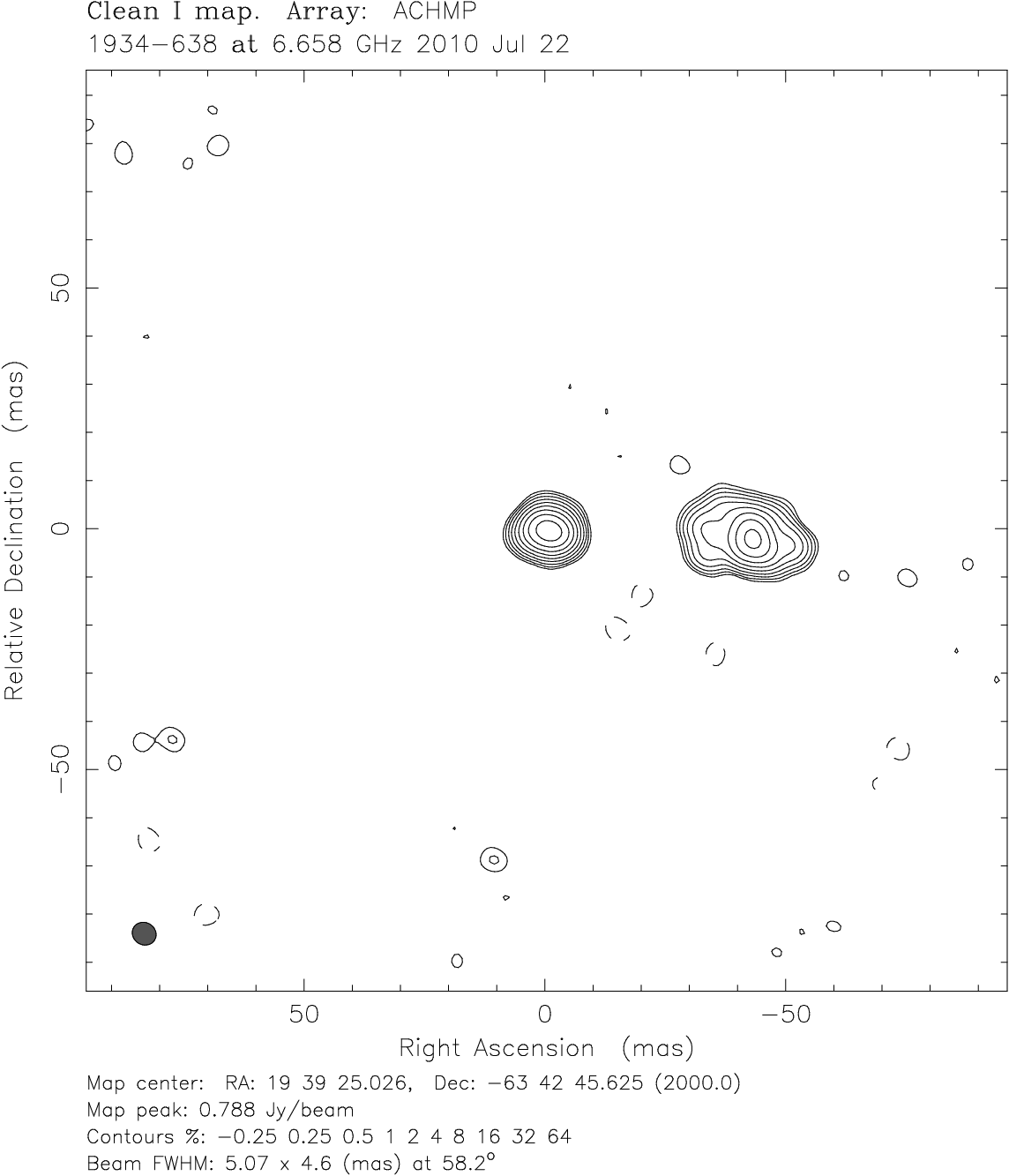}
\end{center}
\caption{The brightness distribution of PKS B1934$-$638 derived from the model created using AIPS and Difmap.}
\label{fig:1934mod}
\end{figure}

\subsection{Initial Calibration\label{sec:calinit}}
The need for amplitude calibration using a calibrator source is well met by PKS B1934$-$638, which has a flux density of 3.983 Jy at 6.7 GHz and does not vary significantly \citep{Reynolds1994b}. Unfortunately, at VLBI resolutions, PKS B1934$-$638 is not a point source and so a model is required in order to make amplitude calibration possible.

\begin{figure}[t]
\begin{center}
\leavevmode
\includegraphics[angle=-90,width=0.4\textwidth]{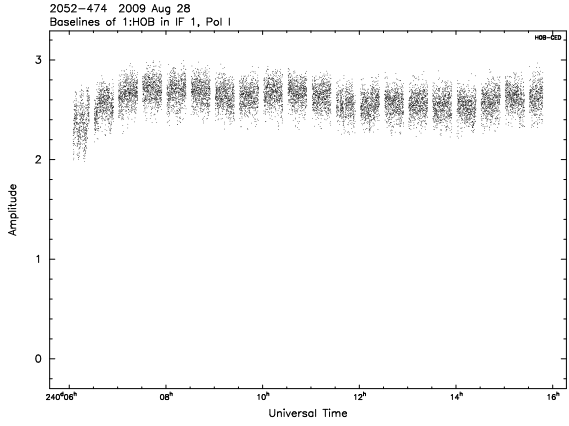}
\end{center}
\caption{Amplitude vs time plot for the calibrated 2052$-$474 observation. Note the downturn in amplitude at the beginning of the experiment. This corresponds to low elevation of the source and is thus likely a residual elevation effect. This is perhaps due to the simple nature of the gain-elevation curve for the Hobart Antenna}
\label{fig:vpl2052}
\end{figure}

To create the model of PKS B1934$-$638, data from the V255 series of Australian LBA observations were used. The V255 project focuses on methanol maser studies at 6.7 GHz, making use of VLBI to measure accurate positions, distances and proper motions of these masers. PKS B1934$-$638 was observed in the V255 observations to act as a polarisation calibrator. Three epochs of V255 were used to make the model of PKS B1934$-$638 used for CHI. As part of the calibration the effect of elevation on the gain of the receiver must be corrected for. These gain-curves are not available for all LBA antennas at 6.7 GHz. Gain-curves from existing 8.4 GHz data were adjusted for the new frequency for telescopes that did not have 6.7 GHz curves available. For Ceduna the gain-curve was taken from \cite{McCulloch2005}. For Hobart it is approximated as a first order polynomial and this may introduce a small elevation effect into CHI data (see Figure \ref{fig:vpl2052}).

Finally the absolute flux scale of the PKS B1934$-$638 model at 6.7 GHz was set using simultaneous ATCA and LBA data of several point sources (0506$-$612, 0808$+$019, 1022$-$665 and 1420$-$679). This was not possible to achieve using the PKS B1934$-$638 data itself as PKS B1934$-$638 is resolved on VLBI baselines. For a point source however the VLBI and ATCA fluxes should be the same and so the difference in VLBI and ATCA flux density was calculated for each of the point sources mentioned above, giving an average correction factor of $1.45 \pm 0.2$ used to scale the model of PKS B1934$-$638. This gave the final clean component model as seen in Figures \ref{fig:1934mod} and \ref{fig:radplsc}. The error in this correction factor (of order 15 percent) is by far the dominant form of error in flux density measurements made using CHI and is thus used as an estimate of the total error in such measurements.

\subsection{CHI Calibration}
As described in section \ref{sec:observations}, PKS B1934$-$638 is observed in each CHI observation to perform additional amplitude calibration, as required by the single baseline. The CHI data (including both the target source(s) and PKS B1934$-$638) are initially calibrated in AIPS \citep[Astronomical Image Processing System;] [] {Greisen1988} in the standard manner. Fringe-fitting is carried out to correct for residual delay and rate errors. This is followed by amplitude calibration using the known gain-curves and system temperature measurements made at each antenna over the course of the experiment.

\begin{figure}[t]
\begin{center}
\leavevmode
\includegraphics[angle=-90,width=0.4\textwidth]{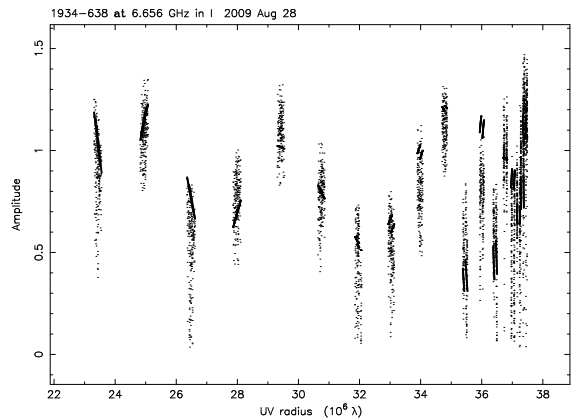}
\end{center}
\caption{Calibrated PKS B1934$-$638 amplitude vs $(u,v)$ distance plot with the overlaid model.}
\label{fig:radplsc}
\end{figure}

 The AIPS calibrated PKS B1934$-$638 data are loaded into the Caltech Difference Mapping program 'Difmap' \citep{Shepherd1995} and the model created as described in Section \ref{sec:calinit} is then overlaid on an amplitude vs $(u,v)$ distance plot. A simple script is then used to manipulate the model to fit both the amplitude vs $(u,v)$ distance plot (see Figure \ref{fig:radplsc}) and an amplitude vs time plot, giving a correction factor. 

This correction factor is then used to scale the visibilities of the baseline producing the final calibrated source observation.

\begin{figure}[htb]
\begin{center}
\leavevmode
\includegraphics[angle=-90,width=0.4\textwidth]{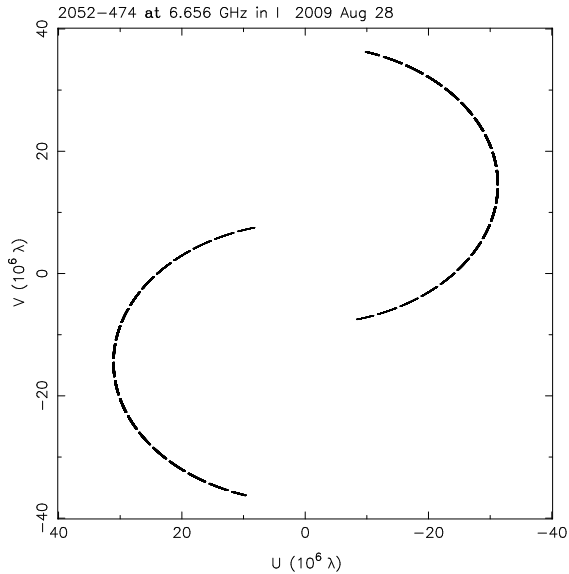}
\end{center}
\caption{The $(u,v)$ coverage obtained from a 12 hour CHI observation of 2052-474. This is typical of a CHI experiment.}
\label{fig:2052_uv}
\end{figure}

\section{Results\label{sec:results}}
We present a typical CHI observation. PKS 2052$-$474 is a blazar at a redshift of 1.489. It is established to be a variable gamma-ray source. \cite{Abdo2010} show a light curve with two flaring events, occurring around February and June 2009 respectively. Previous VLBI observations have found the source to be strongly core dominated on the milli-arcsecond scale \citep{Ojha2010,Ohja2004,Piner2007}. On the kilo-parsec scale, ATCA imaging shows some extended structure \citep{Marshall2005,Burgess2006} extending over ~4 arc-seconds. The observation presented here was made in support of a \textsl{Fermi}/LAT multiwavelength campaign on this object in 2009 \citep{Chang2010} following the flaring seen in the gamma-ray. 

\begin{figure}[htb]
\begin{center}
\leavevmode
\includegraphics[angle=-90,width=0.4\textwidth]{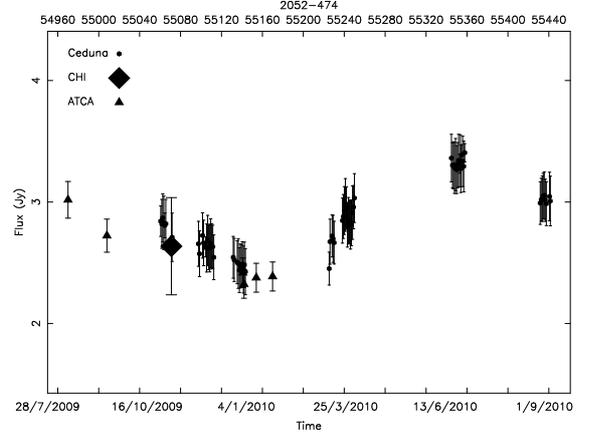}
\end{center}
\caption{Single dish and ATCA monitoring data of PKS 2052$-$474 overlaid with the CHI observation (the large filled diamond). The Ceduna data are the circles and the ATCA data the triangles. The VLBI data is consistent with both the single dish and the ATCA monitoring flux density. Note that the ATCA data are interpolated between simultaneous 5.5 and 9 GHz data to match the 6.7 GHz Ceduna single dish data.}
\label{fig:ced2052}
\end{figure}

\begin{figure}[htb]
\begin{center}
\leavevmode
\includegraphics[angle=-90,width=0.4\textwidth]{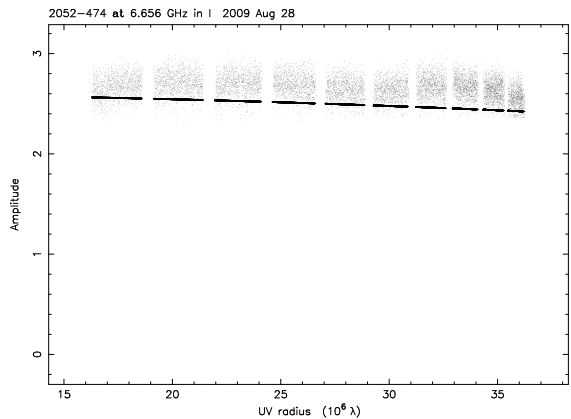}
\end{center}
\caption{Amplitude vs $(u,v)$ distance plot of the calibrated 2052-474 CHI data. A model with major axis of 0.8 mas and an axial ratio of 1 well constrains the visibilities, giving an upper limit on the angular size of the source. Note the choice of model is not a fit to the data but rather lies below the majority of points to give an upper limit to the angular size.}
\label{fig:2052_tb}
\end{figure}

Calibration was performed as explained in section \ref{sec:calibration}, resulting in the amplitude vs time plot seen in Figure \ref{fig:vpl2052} and the amplitude vs $(u,v)$ distance plot seen in Figure \ref{fig:2052_tb}. This is consistent with a point source with a flux density of 2.6 Jy. Figure \ref{fig:ced2052} shows single dish data from the Ceduna monitoring program, as well as ATCA data from the C1730 gamma-ray sources monitoring program, around the time of the CHI observation. The flux density obtained from the CHI observations is consistent with that of the single dish data (as expected for an unresolved source) giving us confidence in the amplitude calibration technique employed by CHI. The $(u,v)$ coverage of a typical CHI observation is shown in Figure \ref{fig:2052_uv}. Applying the radiometer equation (equation \ref{eqn:sens}) to this 7 hour observation of PKS 2052$-$474 gives an expected noise of 0.73 mJy rms, agreeing well with the observed noise in the visibilities of 0.75 mJy rms.

A constraint on the angular size of the source can be calculated by applying a Gaussian model to the visibilities. For this example observation of 2052$-$474 a model with an angular size of 0.8 mas is chosen, as the visibilities are well encompassed by the model (see Figure \ref{fig:2052_tb}). Note that we are not trying to fit to the data directly, rather we are choosing a gaussian which lies below the majority of the visibilities, thus providing an upper limit to the angular size.

The brightness temperature of a source is given by 
\begin{equation}\label{eqn:tb} T_b = \frac{2ln2}{\pi k} \frac{S_\nu \lambda^2 (1+z)}{\theta_{maj}\theta_{min}}\end{equation}

where $k$ is Boltzmann's constant, $S_\nu$ is the flux density of the source, $z$ is its redshift and $\theta$ the angular size (see for example \citealt{Kovalev2005}). 

Using the upper limit to the angular size of the source and equation \ref{eqn:tb}, a lower limit of $2\times10^{11}$ K for the brightness temperature is calculated. This is well below the inverse compton limit.

\section{Summary and Future Developments \label{sec:summary}}
The CHI interferometer is the only facility capable of rapid, high resolution observations of compact radio sources south of about $-30$ degrees declination. The peculiar calibration challenges of such a single baseline interferometer have been met and initial results are presented. Currently CHI is a critical component in the multiwavelength study of AGN in the age of \textsl{Fermi}. However, its role is set to be broader as it is capable of observing any type of compact radio source in the southern sky with high sensitivity. 

There are three developments under active consideration for CHI. First, observation at multiple radio frequencies would provide radio spectral index measurements which are very useful inputs to models of the emission from AGN. Both Hobart and Ceduna can observe at the standard centimeter wavelength radio bands. We will develop the calibration at additional frequencies of 2.3, 8.4 and 22\,GHz, in order to make CHI capable of spectral index measurements. 

Second, while CHI observations can be initiated almost instantly, the remote nature of the Ceduna observing site makes the shipping of data to the correlator in Hobart the biggest bottleneck to rapid analysis. The possibility of taking preliminary measurements using a `sniffing' strategy is being explored. This would involve taking short (about one second) slices of the twelve hours of data and transferring them back via network, thus decreasing the transfer time without sacrificing $(u,v)$ coverage too much. This would allow a quick analysis showing the flaring state of a target (for example) where such information is needed to trigger observations at other wavelengths. 

Finally, we are investigating the possibility of including the new 12 meter antenna at Warkworth, New Zealand \citep{Gulyaev2011} in the CHI array. This would increase the maximum baseline to 3724 kilometers boosting the resolution 2.5 times at an observing frequency of 8.4\,GHz, as well as providing phase closure.


\begin{acknowledgements}
We are extremely grateful to Simon Ellingsen of the University of Tasmania for providing the V255 PKS B1934$-$638 data essential to the proper calibration of CHI. 
This research was funded in part by NASA through \textsl{Fermi} Guest Investigator grant NNH09ZDA001N (proposal number 31263). This research was supported by an appointment to the NASA Postdoctoral Program at the Goddard Space Flight Center, administered by Oak Ridge Associated Universities through a contract with NASA.

This research has made use of data from the NASA/IPAC
Extragalactic Database (NED, operated by the Jet Propulsion Laboratory,
California Institute of Technology, under contract with the National
Aeronautics and Space Administration); and the SIMBAD database (operated
at CDS, Strasbourg, France). This research has made use of NASA's
Astrophysics Data System. This research has made use of the United States
Naval Observatory (USNO) Radio Reference Frame Image Database (RRFID).

\end{acknowledgements}

\bibliographystyle{jwaabib}
\bibliography{mnemonic,aa_abbrv,tanami}

\begin{thebibliography}{}

\bibitem[\protect\astroncite{{Abdo} et~al.}{2010}]{Abdo2010}
{Abdo} A.A., {Ackermann} M., {Ajello} M., et~al., 2010, \apj 722, 520

\bibitem[\protect\astroncite{{Agudo} et~al.}{2011}]{Agudo2011}
{Agudo} I., {Jorstad} S.G., {Marscher} A.P., et~al., 2011, \apjl 726, L13+

\bibitem[\protect\astroncite{{Atwood} et~al.}{2009}]{Atwood2009}
{Atwood} W.B., {Abdo} A.A., {Ackermann} M., et~al., 2009, ApJ 697, 1071

\bibitem[\protect\astroncite{{Briggs}}{1983}]{Briggs1983}
{Briggs} F.H.,  1983, \aj 88, 239

\bibitem[\protect\astroncite{{Burgess} \& {Hunstead}}{2006}]{Burgess2006}
{Burgess} A.M., {Hunstead} R.W.,  2006, \aj 131, 114

\bibitem[\protect\astroncite{{Chang} et~al.}{2010}]{Chang2010}
{Chang} C.S., {Ros} E., {Kadler} M., et~al., 2010, ArXiv e-prints 1001.1563

\bibitem[\protect\astroncite{{Cohen} et~al.}{2007}]{Cohen2007}
{Cohen} M.H., {Lister} M.L., {Homan} D.C., et~al., 2007, ApJ 658, 232

\bibitem[\protect\astroncite{{Deller} et~al.}{2007}]{Deller2007}
{Deller} A.T., {Tingay} S.J., {Bailes} M., {West} C.,  2007, PASP 119, 318

\bibitem[\protect\astroncite{{Greisen}}{1998}]{Greisen1988}
{Greisen} E.W.,  1998,
\newblock In: {Albrecht} R., {Hook} R.N., {Bushouse} H.A. (eds.) Astronomical
  Data Analysis Software and Systems VII, Vol. 145. Astronomical Society of the
  Pacific Conference Series, p.204

\bibitem[\protect\astroncite{{Gulyaev} et~al.}{2011}]{Gulyaev2011}
{Gulyaev} S., {Natusch} T., {Weston} S., et~al., 2011, ArXiv e-prints 1103.2830

\bibitem[\protect\astroncite{{Jorstad} et~al.}{2001}]{Jorstad2001}
{Jorstad} S.G., {Marscher} A.P., {Mattox} J.R., et~al., 2001, \apj 556, 738

\bibitem[\protect\astroncite{{Kovalev} et~al.}{2005}]{Kovalev2005}
{Kovalev} Y.Y., {Kellermann} K.I., {Lister} M.L., et~al., 2005, AJ 130, 2473

\bibitem[\protect\astroncite{{Lovell} et~al.}{2003}]{Lovell2003}
{Lovell} J.E.J., {Jauncey} D.L., {Bignall} H.E., et~al., 2003, AJ 126, 1699

\bibitem[\protect\astroncite{{Lovell} et~al.}{2008}]{Lovell2008}
{Lovell} J.E.J., {Rickett} B.J., {Macquart} J.P., et~al., 2008, \apj 689, 108

\bibitem[\protect\astroncite{{Marshall} et~al.}{2005}]{Marshall2005}
{Marshall} H.L., {Schwartz} D.A., {Lovell} J.E.J., et~al., 2005, ApJS 156, 13

\bibitem[\protect\astroncite{{McCulloch} et~al.}{2005}]{McCulloch2005}
{McCulloch} P.M., {Ellingsen} S.P., {Jauncey} D.L., et~al., 2005, AJ 129, 2034

\bibitem[\protect\astroncite{{M{\"u}ller} et~al.}{2011}]{Mueller2011}
{M{\"u}ller} C., {Kadler} M., {Ojha} R., et~al., 2011, \aap 530, L11+

\bibitem[\protect\astroncite{{Ojha} et~al.}{2004a}]{Ojha2004}
{Ojha} R., {Fey} A.L., {Johnston} K.J., et~al., 2004a, AJ 127, 3609

\bibitem[\protect\astroncite{{Ojha} et~al.}{2004b}]{Ohja2004}
{Ojha} R., {Fey} A.L., {Johnston} K.J., et~al., 2004b, \aj 127, 3609

\bibitem[\protect\astroncite{{Ojha} et~al.}{2010}]{Ojha2010}
{Ojha} R., {Kadler} M., {B{\"o}ck} M., et~al., 2010, \aap 519, A45+

\bibitem[\protect\astroncite{{Pearson} \& {Readhead}}{1984}]{Pearson1984}
{Pearson} T.J., {Readhead} A.C.S.,  1984, \araa 22, 97

\bibitem[\protect\astroncite{{Phillips} et~al.}{2009}]{Phillips2009}
{Phillips} C., {Tzioumis} T., {Tingay} S., et~al., 2009,
\newblock In: 8th International e-VLBI Workshop.

\bibitem[\protect\astroncite{{Piner} et~al.}{2007}]{Piner2007}
{Piner} B.G., {Mahmud} M., {Fey} A.L., {Gospodinova} K.,  2007, \aj 133, 2357

\bibitem[\protect\astroncite{{Readhead} et~al.}{1980}]{Readhead1980}
{Readhead} A.C.S., {Walker} R.C., {Pearson} T.J., {Cohen} M.H.,  1980, \nat
  285, 137

\bibitem[\protect\astroncite{{Reynolds}}{1994}]{Reynolds1994b}
{Reynolds} J.E.,  1994,
\newblock A revised flux scale for the AT Compact Array,
\newblock ATNF Memo AT/39.3/040

\bibitem[\protect\astroncite{{Schl{\"u}ter} \& {Behrend}}{2007}]{Schluter2007}
{Schl{\"u}ter} W., {Behrend} D.,  2007, Journal of Geodesy 81, 379

\bibitem[\protect\astroncite{{Shepherd} et~al.}{1995}]{Shepherd1995}
{Shepherd} M.C., {Pearson} T.J., {Taylor} G.B.,  1995,
\newblock In: {B.~J.~Butler \& D.~O.~Muhleman} (ed.) Bulletin of the American
  Astronomical Society, Vol. 27. Bulletin of the American Astronomical Society,
  p.903

\bibitem[\protect\astroncite{Thompson et~al.}{2001}]{thompson2001}
Thompson A., Moran J., Swenson G.,  2001,
\newblock Interferometry and synthesis in radio astronomy,
\newblock Wiley

\bibitem[\protect\astroncite{{Ulrich} et~al.}{1997}]{Ulrich1997}
{Ulrich} M.H., {Maraschi} L., {Urry} C.M.,  1997, \araa 35, 445

\bibitem[\protect\astroncite{{Wagner} \& {Witzel}}{1995}]{Wagner1995}
{Wagner} S.J., {Witzel} A.,  1995, \araa 33, 163

\bibitem[\protect\astroncite{{Walker}}{1995}]{Walker1995}
{Walker} R.C.,  1995,
\newblock In: {J.~A.~Zensus, P.~J.~Diamond, \& P.~J.~Napier} (ed.) Very Long
  Baseline Interferometry and the VLBA, Vol. 82. Astronomical Society of the
  Pacific Conference Series, p.133

\end{thebibliography}

\clearpage

\begin{deluxetable}{lrr}
\tabletypesize{\scriptsize}
\tablecaption{Brightness Temperature Sensitivity\label{table:brightnesstemp}}
\tablewidth{0pt}
\tablehead{
\colhead{Redshift} & \colhead{Shape} & \colhead{Brightness Temperature\tablenotemark{a}} \\
\colhead{} & \colhead{}  & \colhead{K}
}
\startdata
0 & Gaussian & $1\times10^{9}$ \\
0 & Sphere & $0.5\times10^{9}$ \\
0 & Disc  & $0.4\times10^{9}$\\
1 & Gaussian & $2.2\times10^{9}$ \\
1 & Sphere & $1\times10^{9}$ \\
1 & Disc  & $0.8\times10^{9}$\\
\enddata
\tablenotetext{a}{For a 1 jansky source.}
\end{deluxetable}

\end{document}